# Microstructural magnetic phases in superconducting FeTe$_{0.65}$Se$_{0.35}$

A. Wittlin, P. Aleshkevych, H. Przybylińska, D. J. Gawryluk, P. Dłużewski, M. Berkowski, R. Puźniak, M. U. Gutowska and A. Wiśniewski

*Institute of Physics, Polish Academy of Sciences, Aleja Lotników 32/46, PL-02-668 Warsaw, Poland*

E-mail: wittlin@ifpan.edu.pl

**Abstract**
In this paper, we address a number of outstanding issues concerning the nature and the role of magnetic inhomogenities in the iron chalcogenide system FeTe$_{1-x}$Se$_x$ and their correlation with superconductivity in this system. We report morphology of superconducting single crystals of FeTe$_{0.65}$Se$_{0.35}$ studied with transmission electron microscopy, high angle annular dark field scanning transmission electron microscopy and their magnetic and superconducting properties characterized with magnetization, specific heat and magnetic resonance spectroscopy. Our data demonstrate a presence of nanometre scale hexagonal regions coexisting with tetragonal host lattice, a chemical disorder demonstrating non homogeneous distribution of host atoms in the crystal lattice, as well as hundreds-of-nanometres-long iron-deficient bands. From magnetic data and ferromagnetic resonance temperature dependence, we attribute magnetic phases in Fe-Te-Se to Fe$_3$O$_4$ inclusions and to hexagonal symmetry nanometre scale regions with structure of Fe$_7$Se$_8$ type. Our results suggest that nonhomogeneous distribution of host atoms might be an intrinsic feature of superconducting Fe-Te-Se chalcogenides and we find a surprising correlation indicating that faster grown crystal of inferior crystallographic properties is a better superconductor.

## 1. Introduction

Superconductivity in layered iron chalcogenides ("11"-type system compounds) is currently the subject of intensive research, covering the search for the mechanism of superconductivity, efforts to increase the critical temperature and the understanding of the interplay between superconductivity and magnetism. The critical temperature $T_c$ of the parent compound $\beta$-FeSe is ~ 8 K and it rises with applied hydrostatic or chemical pressure. Substitution of Se with Te increases $T_c$ up to ~ 15 K and upon intercalating of "11" with potassium, rubidium, or cesium $T_c$ is further increased up to ~ 30 K. FeSe has a simple chemical formula and structurally it is one of the simplest compounds of the recently discovered iron-based high $T_c$ families. As such, it has attracted considerable attention worldwide; see for example [1-4]. Moreover, the high upper critical magnetic field of this system makes it a promising material for applications in new types of high-field superconducting wires [5] and tapes [6].

In FeTe$_{1-x}$Se$_x$, the square lattice of Fe atoms is tetrahedrally coordinated by chalcogene ions and its structure has no dedicated charge reservoir layer characteristic for iron pnictides. The magnetic phase diagram contains four distinct phases [7-14]. It is a tetragonal paramagnetic metal for all $x$ at high temperatures, it shows spin-density wave ordering for $x < 0.1$, spin glass-like static magnetic ordering for $0.1 < x < 0.3$ and superconductivity for $x > 0.3$ at low temperatures. Although FeTe$_{1-x}$Se$_x$ appears to be an almost ideal model system for the study of the phenomenon of superconductivity in iron based compounds, the detailed analysis of data is significantly hindered by an intrinsic and extrinsic crystal disorder which results from a complex structural chemistry and an apparent inherent non-stoichiometry. Multi-scale lattice disorder begins at short range atomic level in the mixed crystal because the Te and Se ions are at slightly different positions in the unit cell [15,16]. On larger distance scales, crystals of FeTe$_{1-x}$Se$_x$ tend to have Fe non-stoichiometry often described as the Fe$_7$Se$_8$ type, non-homogeneities like clustering and microstructural foreign phases of Fe chalcogenides [12,17-20]. Since some of these phases have distinct magnetic properties they also mask and distort the intrinsic response of the parent compound. Understanding of these phenomena appears to be essential for the elucidation of the underlying mechanism of superconductivity in this system.

In this contribution, we present morphology, magnetization, specific heat and magnetic resonance (MR) spectroscopy results obtained for two different single crystals of the FeTe$_{0.65}$Se$_{0.35}$ system. The main observation of our study is that the crystals of apparently inferior quality exhibit more pronounced superconductivity and sharper superconducting transition.

Our Te-rich superconducting FeTe$_{0.65}$Se$_{0.35}$ single crystals, grown with two different velocities in two procedurally identical growth processes, have identical nominal composition and the same measured $T_c$ (onset) at ~ 12.9 K. Nevertheless, these two samples exhibit different crystallographic and physical, normal and superconducting, properties. It is found that the sharpness of transition to the superconducting state is strongly correlated with the crystallographic quality of the studied crystals. The $\Delta\omega$ value, describing the full width at half maximum (FWHM) of the 004 x-ray diffraction peak, obtained in the $\omega$ scan measurements was chosen as a criterion of crystallographic quality since changes in the $c$-axis lattice constant are very sensitive to the variation in chemical composition of studied materials [21], the 004 peak is relatively intense and appears at sufficiently large angles to get good angular resolution.

## 2. Samples preparation

Superconducting single crystals of FeTe$_{0.65}$Se$_{0.35}$ have been grown using Bridgman's method. The samples were prepared from stoichiometric quantities of iron chips (3N5), tellurium powder (4N) and selenium powder (pure). Double-walled evacuated sealed quartz ampoules containing the starting materials were placed in a furnace with a vertical gradient of temperature regulated from 0.4 to 2 °C/mm. The samples were synthesized for 6 h at a temperature of 680 °C which was then raised to 920 °C. After sample melting, the temperature was held for 3 h and then reduced down to 400 °C at the rate of 1 – 4 °C/h, then to 200 °C at the rate of 60 °C/h and finally to room temperature. Proper adjustment of cooling velocity and/or vertical gradient of temperature in the furnace allowed us to tune the growth velocity in a range from ~ 0.5 to ~ 10 mm/h, permitting the growth of single crystals of various crystallographic quality. The obtained crystals exhibited a (001) cleavage plane with a random orientation with respect to the growth direction. Details of the samples' structural (x-ray) and chemical (scanning electron microscopy/energy dispersive x-ray spectroscopy – SEM/EDX) analysis have been published elsewhere [21]. For our study, we have selected two crystals obtained from two different growth processes, henceforth labelled A and B, with different crystal quality. The crystals, with $\Delta\omega$ values of the 004 x-ray diffraction peak equal to 6.00 (crystal A) and 1.67 arc min (crystal B), have been grown with velocities of ~ 8 and ~ 1.2 mm/h, respectively (see the table 1).

## 3. Experimental details

X-ray powder diffraction (XRPD) of the crystals was performed at room temperature using Siemens D5000 diffractometer with Ni-filtered Cu K$\alpha$ radiation. Data on powdered single crystals were collected in the angular range $25° < 2\theta < 95°$ with step of 0.02° and an averaging time of 10 s/step. The diffraction patterns were analyzed by the Rietveld refinement method using DBWS-9807 program [22]. Accurate values of the $c$ lattice constant and the $\Delta\omega$ value of the $\omega$ scan on the 004 diffraction line were obtained in single-crystal measurements on the well defined, natural cleavage (00l) plane. The $c$ lattice constant obtained in single-crystal measurements was used as a fixed value in the powder Rietveld analysis for the determination of other structural parameters, i.e. $a$, $V$ and the occupation number.

The chemical composition of the matrix and inclusions was checked on the cleavage plane of the crystals by field emission scanning electron microscopy (FESEM) JEOL JSM-7600F operating at 20 kV. The quantitative point analyses were performed by Oxford INCA energy dispersive x-ray spectroscopy (EDX) coupled with the SEM. The specimens for transition electron microscopy (TEM) were prepared by Ar ion milling in a PIPS Gatan device and examined with the use of a Titan Cubed 80–300 Cs corrected microscope.

The measurements of AC magnetic susceptibility (field amplitude 1 Oe and 10 Oe, frequency 10 kHz) were performed with a Physical Property Measurement System (PPMS) and DC magnetization with Magnetic Property Measurement System (MPMS) of Quantum Design. Magnetic resonance spectra were collected using a commercial continuous wave x-band Bruker E580 spectrometer equipped with continuous flow LHe cryostat for cooling the sample over the temperature range of 2 – 300 K.

## 4. Results and discussion

Figure 1 shows x-ray powder diffraction patterns for crystals A and B performed: just after growth process (A$_0$, B$_0$), 20 months after grown (A$_1$ - from the same ingot as A$_0$) and 14 months after grown (B$_1$ - from the same ingot as B$_0$). Major phase reflections were indexed to a tetragonal cell in the space group *P4/nmm* (No. 129) of the PbO structural type with occupation of Wyckoff's *2a* site by Fe and the *2c* site by Se/Te. It was assumed that excess Fe ions occupy *2c* site of structural vacancy in the Se/Te plane [23,24]. Additional Bragg peaks could be indexed to the hexagonal phase of Fe$_7$Se$_8$ type or to iron oxides inclusions. The additional, faint, peaks in XRD patterns A$_1$ and B$_1$ collected after a long period of time after crystal growth are not well understood. In fact, a recent Mössbauer study [17] did not show any oxidation or aging effect in Fe-Te-Se and even prolonged exposition of crystals to air did not bring an increase of the amount of Fe$_3$O$_4$. Moreover, we did not observe any changes in superconducting properties of crystals over time, such as were reported, for example, for FeTe$_{0.8}$S$_{0.2}$ [25].

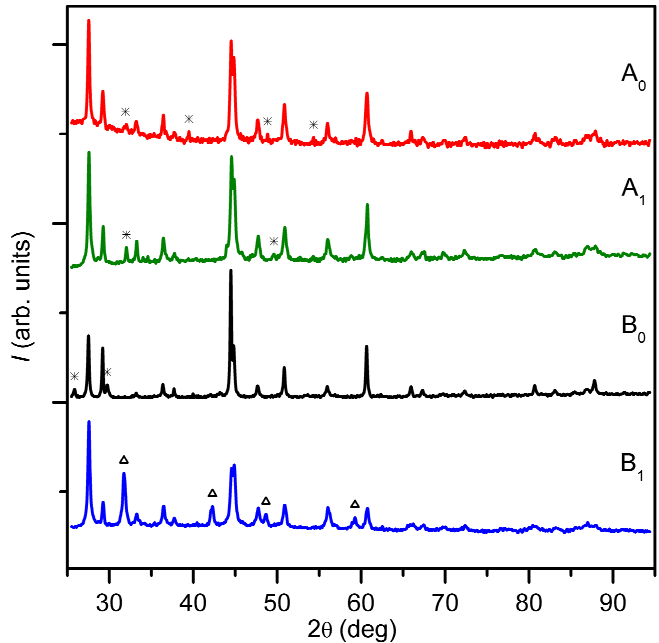

**Figure 1.** Time evolution of x-ray powder diffraction (XRPD) (using Cu K$\alpha$ radiation) patterns for samples A (A$_0$, A$_1$) and B (B$_0$, B$_1$) performed: just after grown (A$_0$, B$_0$), 20 months after grown (A$_1$) and 14 months after grown (B$_1$). Major phase reflections were indexed to a tetragonal cell in the space group *P4/nmm* (No. 129) of the PbO structural type. Additional Bragg reflexes (marked by asterisk) may be attributed to iron oxides inclusions and additional Bragg peaks in B$_1$ (marked by triangles) originate most likely from the hexagonal phase. The $R_p$ form factor, obtained by the Rietveld refinement method using DBWS-9807 program, is equal to 0.037 ($N = 4700$), 0.033 ($N = 3457$), 0.057 ($N = 3329$) and 0.034 ($N = 3502$) for A$_0$, A$_1$, B$_0$, B$_1$, respectively (where *N* is a number of analyzed points).

Figure 2 shows typical TEM images of a 20 × 20 nanometres *a-b* plane surface of sample A and B specimens as selected from larger crystalline samples. Submicron resolution TEM data show in both A and B samples long range ordered tetragonal matrix with more or less uniformly distributed regions of hexagonal symmetry. The hexagonal regions have approximately oval shapes, forming strips and are surrounded by an intermediate region of less pronounced symmetry. High-resolution transmission electron images show a coexistence of the tetragonal (T) and hexagonal (H) phases with the same orientation of their axes. In the <001> zone axis, high-resolution images the {100} planes remain almost the same for tetragonal, intermediate and hexagonal regions. Only the inter-planar distance between {100} planes becomes different according to the lattice constant $a_H : a_T$ ratio. Our data demonstrate that these hexagonal regions introduce tensile stress along the tetragonal strip direction. In both A and B samples, the tetragonal strips are 10 to 20 nm wide and hundreds of nanometers long. The TEM data of figure 2 also show a pronounced difference in the morphology of both samples. In the sample A, hexagonal regions have a typical diameter of 1 to 3 nm, while in the sample B we observe distinct hexagonal regions with diameter of 10 nm and larger. These iron deficient regions show NiAs type symmetry, henceforth described as Fe$_7$(Te,Se)$_8$ phase.

The observed tendency to organize iron vacancies into clusters during crystal growth process leads to creation of hexagonal structure regions in the tetragonal host matrix. The size and distribution of such regions depend on the speed of crystal growth. In crystals grown with high growth velocity, there are many such iron vacancy regions, which are small and randomly distributed. Therefore, these hexagonal regions disturb the host tetragonal matrix, inducing local disorder and local strains, which manifest in broadening of the (004) x-ray diffraction peak. On the other hand, in crystals grown slowly iron vacancies form large clusters, while non-disturbed host tetragonal matrix is well ordered on much larger distances. It is known from the literature [1,26] that in-growth Fe$_7$(Te,Se)$_8$ phase regions are often present in FeTeSe crystals and, as will be shown below, their signature can also be observed in the magnetic response of both samples.

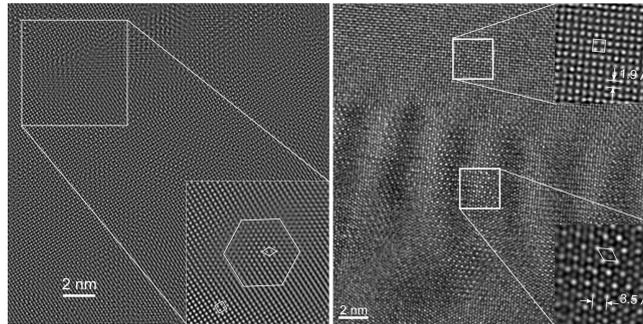

**Figure 2.** The sample A (left panel) – tetragonal strip and nanometre size hexagonal region (projections along the *c*-axis of hexagonal NiAs type and tetragonal unit cells) and intermediate region (pointed by a hexagon). The sample B (right panel) – the tetragonal band (upper part) and larger than 10 nm size regions with NiAs type structure (bottom part). Left bottom: the enlargement of white rectangle presents the tetragonal and hexagonal structure with the projection of unit cells. The (200) and (100) interplanar distances for the tetragonal and hexagonal structures respectively are shown in the right panel.

Figure 3 shows pictures of the *Z*-contrast imaging of larger, 600 × 600 nanometres, samples of A and B as generated by high angle annular dark field (HAADF) scanning transmission electron microscopy. This technique is ideal for tomographic imaging of thin crystalline samples. As shown in figure 3, it generates strong contrast with an intensity which is approximately proportional to $Z^2$ (where *Z* is the atomic number).

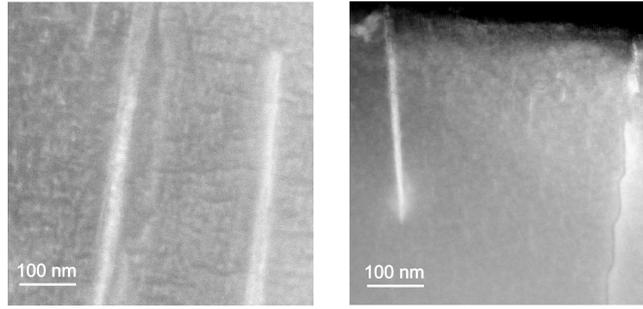

**Figure 3.** STEM HAADF images of the sample A (left panel) and B (right panel) – the EDX measurements indicate that the white bands have lower (of few at%) concentration of Fe. The left image shows fine structure of the region between the strips as opposed to the right image where the contrast seems rather uniform. The white bands on the left image are more diffused than on the right one.

The main features of these pictures are the micrometers long and about 10 nm wide white bands. The energy dispersive x-ray (EDX) spectroscopy measurements of Fe, Se and Te distribution lead us to the conclusion that inside these bands the iron concentration is lower by few atomic percent than in the adjacent sample volume. These bands are broad and diffused in the sample A and narrow and sharp in the sample B. In addition, the background image of the sample A display a non-uniform pattern of $Z$-contrast with nanometre size scale. In the sample B, on the contrary, the image background $Z$-contrast is rather uniform. A possible origin of such $Z$-contrast image pattern in the sample A, displaying apparent chemical inhomogeneity, will be discussed below. In conclusion, HRTEM observations indicate that the structure of the sample B is different from that for the sample A by a presence of hexagonal regions of 10 nm and larger diameter and at the same time the absence of a nanometre-scale non-uniform charge modulation. These results are consistent with $\Delta\omega$ x-ray measurements.

Since the investigated samples have different shapes, the impact of demagnetizing field on the AC susceptibility varies between them. For comparison of the sharpness of superconducting state transition, the data for $4\pi\chi'$ were, therefore, normalized to the value of **-1** for real part of AC susceptibility at low temperatures. The same procedure was used to normalize data of the imaginary part $4\pi\chi''$. Figure 4 shows the temperature dependence of the bulk magnetic susceptibility and of the specific heat near $T_c$. Both samples, A and B, clearly show almost the same onset $T_c$ in the temperature dependence of the real part of the AC susceptibility $\chi'(T)$. However, the shape of the susceptibility curve and the temperature dependence of the specific heat for both samples are qualitatively and quantitatively different.

Sample A data exhibit a typical "good superconductor" behaviour with a narrow transition width in $\chi'(T)$, as shown in figure 4 b and a distinct specific heat anomaly at $T_c = 12.5$ K, clearly visible in figure 4 c. On the other hand, the transition width of $\chi'(T)$ for the sample B is very broad and it extends over 10 K, down to 2 K. The specific heat anomaly near $T_c$ for the sample B is small and it is not easily visible when the data for both samples are plotted on the same scale, as in figure 4 c. In that picture, the sample B specific heat temperature dependence is almost featureless near $T_c$ and the data can be quite well approximated by a textbook dependence: $\gamma T + BT^3 + CT^5$, with the electronic ($\gamma T$) and lattice ($BT^3 + CT^5$) contributions to the specific heat only, with $\gamma = 45$ mJ/(mol K$^2$) and $\Theta_D = 184$ K (the Debye temperature $\Theta_D$ is related to specific heat through

$$C_p = C_V + R = \frac{12\pi^4}{5} N_A k_B \left(\frac{T}{\Theta_D}\right)^3 + R$$

). These values are similar to those reported in a recent study [27].

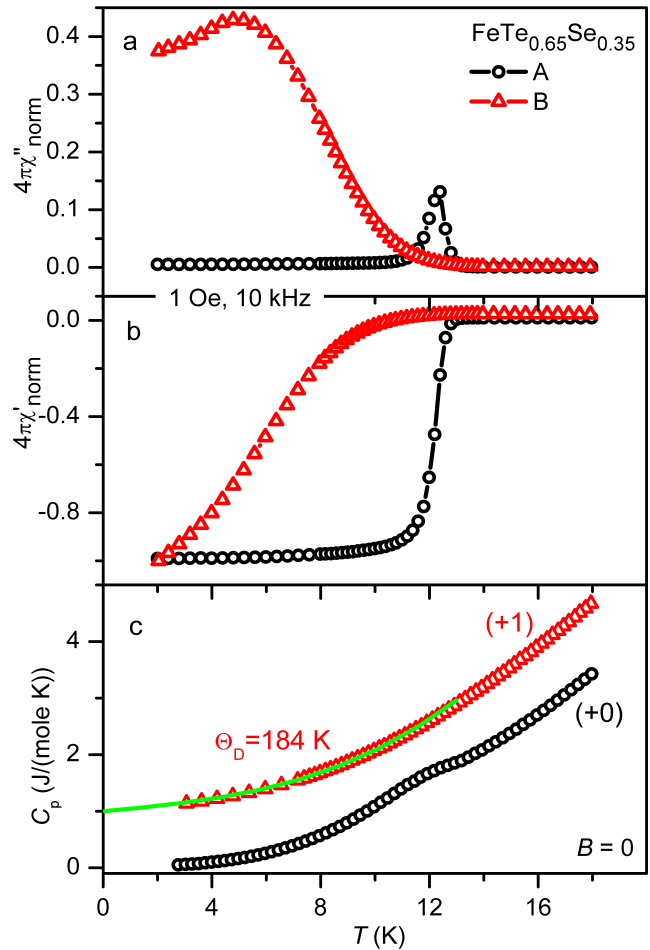

**Figure 4.** Temperature dependence of the imaginary part (a) and the real part (b) of AC magnetic susceptibility, normalized to the ideal value of −1 for the real part of AC susceptibility, measured in 1 Oe of AC field with 10 kHz in warming mode and specific heat (c) for two single crystals of FeTe$_{0.65}$Se$_{0.35}$ (curve for the sample B is shifted up along $C_p$ axis by the value +1). The solid curve in (c) shows fitted lattice and electronic (incoherent) contributions to the specific heat.

Figure 5 shows the temperature dependence of the real part of magnetic susceptibility recorded in magnetic field of 10 Oe (figure 5 a) and of magnetization recorded in magnetic field of 10 kOe applied parallel to the *c*-axis of the crystal (figure 5 b) for both samples A and B over the temperature range between 300 K and $T_c$. Temperature dependence of the magnetization recorded for single crystal A is very similar to that of FeSe$_{0.5}$Te$_{0.5}$, recorded in a magnetic field of 10 kOe, applied parallel to the *c*-axis of the crystal and attributed to the existence of an impurity phase of Fe$_7$(Te,Se)$_8$ [27]. Importantly, magnetic data recorded for both samples show a distinct feature at around 125 K. Fe$_7$Se$_8$ is known to undergo a spin-axis transition below 130 K leading to a reduction in magnetization for *H* parallel to the *c*-axis [28,29], as observed in the studied sample. On the other hand, the change in the temperature dependence of AC susceptibility observed at approx. 125 K may be attributed also to the Verwey transition [30,31] in Fe$_3$O$_4$. The value of saturation magnetization, $M_s$, for the studied crystal is of about 0.86 emu/g at 80 K. Saturation magnetization of Fe$_7$Se$_8$, determined by Kamimura [29] at 80 K, is of about 85 emu/cm$^3$, what corresponds to about 14 emu/g. Those values of saturation magnetization lead to an estimation of the maximum volume fraction of Fe$_7$(Te,Se)$_8$ phase in the studied samples not exceeding 6%. It correlates well with 5.35(40)% estimation of the volume fraction for impurity hexagonal Fe$_7$(Te,Se)$_8$ phase (space group *P6$_3$/mmc*) in FeTe$_{0.5}$Se$_{0.5}$ obtained from neutron powder diffraction measurements performed on a similar crystal [14]. Saturation magnetization of Fe$_3$O$_4$ nanoparticles depends on their size and decreases with decreasing nanoparticle size. For the smallest nanoparticles with the size of about 5 nm, the magnitude of the magnetization at 20 K in magnetic field of 10 kOe is larger than 40 emu/g [32]. It means that volume fraction of the impurity Fe$_3$O$_4$ phase in the studied sample does not exceed 2%. Powder x-ray data for the studied samples performed just after growth processes give upper estimation of iron oxides impurities volume fraction to 5%. The susceptibility data show that the transition at ~ 120 K is sharper for the sample A and a similar conclusion can be drawn from magnetization data. In particular, the sample A shows a much more distinct hysteretic behaviour below 120 K. Furthermore, there is stronger paramagnetic-like background for the sample B than for the sample A. Obtained data indicate that both studied samples contain almost equal volume fraction of magnetic impurity phase or phases. However, the impurity phase/phases in the sample A is/are much better developed and separated from the parent compound than that/those in the sample B. In the existing literature, magnetic anomaly (phase transition) at ~ 120 K has been discussed in many reports on the Fe-Te-Se system [17,19,33]. Several authors [20,34], have also seen it in electrical [35] or thermal [34,36] transport measurements.

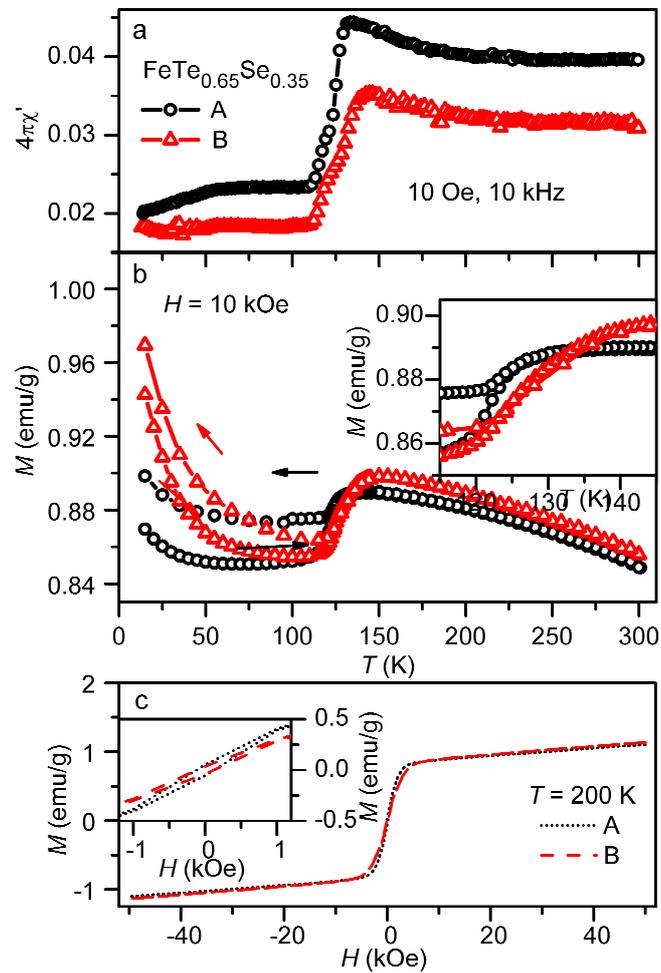

**Figure 5.** (a) Temperature dependence of the real part of AC magnetic susceptibility for two single crystals of FeTe$_{0.65}$Se$_{0.35}$ in 10 Oe of AC field with 10 kHz measured in the temperature range above $T_c$ in the warming mode. (b) Temperature dependence of DC magnetization in 10 kOe recorded after zero field cooling in the temperature range above $T_c$ in warming and cooling modes. (c) Field dependence of the magnetization in a magnetic field up to 50 kOe at 200 K. Inset to panel (b): expansion of $M(T)$ data in 10 kOe for the temperature range between 115 and 145 K; inset to panel (c): expansion of $M(H)$ data between –1.2 and +1.2 kOe at 200 K.

To establish the origin of magnetization feature observed at ~ 120 K in our crystals we did a comprehensive study of the angular and temperature dependencies of magnetic resonance (MR) spectra for comparison with the structural study and magnetic properties shown above.

Only a few results of MR studies have been reported on iron-based "11" superconducting crystals until now. Recently, Arcon *et al* [37] in their studies of x-band electron paramagnetic resonance (EPR) on FeTe$_{0.58}$Se$_{0.42}$, performed between 400 K and 100 K, found a single line Dyson-shape spectrum, with a strong angular dependence of the line width. They observed a broadening of the spectrum and an increase of *g* factor with decreasing temperature and concluded that the line originates from both quasi-particle states (charge carriers) and unidentified localized states. Li *et al* [38] reported EPR results for an intercalated K$_{0.8}$Fe$_2$Se$_{1.4}$S$_{0.4}$ superconducting single crystal at the temperature range from 2 K to 300 K. They also observed an asymmetric temperature-dependent spectrum which broadens with decreasing temperature. That spectral feature disappears below 70 K and it is assigned to local moments of Fe ions. Moreover, the authors observed an additional, unexplained, spectral feature at low magnetic fields at temperatures below 20 K, attributed either to superconductivity-enhanced local moments of Fe ions or to "a novel magnetic state".

Our magnetic resonance spectra were collected on plate-shaped crystalline samples selected from cleaved pieces from the bulk material. Many such samples have been investigated and the representative spectra for the crystals A and B are shown in figure 6. The spectra were recorded at room temperature with magnetic field oriented in the *c*-plane of the crystals. In both crystals, two broad, poorly resolved, asymmetric resonances are observed, with the resonance fields varying slightly with crystal orientation in respect to the external magnetic field applied. Their apparent line shape depends on the actual specimen investigated.

The MR spectra which we attribute, as discussed below, to the collective magnetic excitation, ferromagnetic resonance (FMR), were analyzed with use of a standard method [39] suitable for broad resonance lines, which takes into account that linearly polarized microwave field in the resonant cavity contains both clockwise and counterclockwise polarization directions, affecting the measured line shape. The single resonance line shape is described [39] by:

$$\frac{dP}{dH} = \frac{1}{2}\frac{d}{dH}\left(\chi_+'' + \chi_-''\right), \qquad (1)$$

where *P* is the absorbed microwave power and $\chi_\pm''$ are Lorentzian line profiles for +/- polarization of the microwave field [40] given by:

$$\chi_\pm'' = \frac{2I}{\pi}\frac{\Delta H}{4(H \mp H_0)^2 + \Delta H^2}. \qquad (2)$$

The parameter *I* is the integral intensity, *ΔH* is the line width at half-maximum and $H_0$ is the resonance field. The spectra in figure 6 were fitted with two different Lorentzians (given by equations 1, 2) for each sample. The best fit was obtained with parameters given in table 2 and the resolved individual Lorentzians (labeled (1), (2) for the crystal A and (3), (4) for the crystal B) are displayed with dashed lines in figure 6.

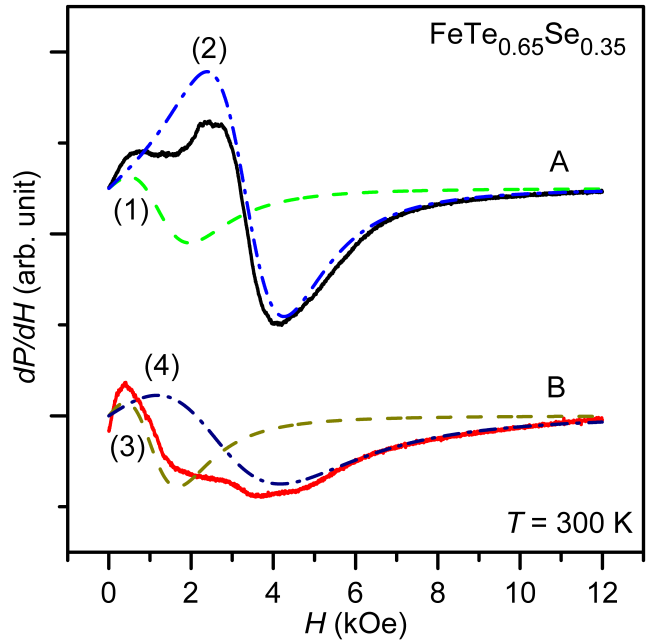

**Figure 6.** MR spectra for the A and B samples recorded with $H \perp c$-axis at room temperature. The dashed lines show fits to the main spectral features observed (see text).

The most intense resonances in the crystals A and B (lines (2) and (4), respectively) show a similar, weak dependence of the resonance field position $H_0$ on the orientation of the magnetic field $H$. The polar ($H$ rotated by $\theta$ about an in-plane crystal axis) and azimuthal ($H$ rotated by $\varphi$ about the $c$-axis) angular dependencies of $H_0$ for line (2) at 300 K are depicted in figure 7 a. Clearly visible is the 6-fold symmetry in the $c$-plane (the solid line in the lower panel of figure 7 a visualizes a $\cos(6\varphi)$ dependence). In the upper panel of figure 7 a contributions of $\sin^2\theta$, $\sin^4\theta$ and $\sin^6\theta$ dependencies can be detected. Such a variation of the resonance fields with angle is not expected for a tetragonal symmetry lattice. In contrast, it is characteristic of magnetocrystalline anisotropy with hexagonal symmetry. Thus, the observed resonances dominating typically in the FMR spectra of both crystals A and B must stem from inclusions of a different crystallographic phase than the host lattice. Moreover, such distinct magnetocrystalline anisotropy can be only observed if all the inclusions are identically oriented crystallographically. The broadness of the resonance lines, however, points out to a considerable shape distribution of the particles [41]. The larger line widths typically encountered in specimens of the crystal B (compare $\Delta H$ for lines (2) and (4) in table 2) indicate, moreover, a less uniform shape distribution than in the crystal A.

In a few specimens of the crystal B, in addition to the lines (3) and (4), a broad and extremely asymmetric resonance line was also detected. The microwave absorption spectrum recorded for one sample, where this resonance was dominant, is shown in the lower panel of figure 7 b for two orientations of magnetic field ($H \parallel c$ and $H \perp c$). This resonance is characterized by a very strong angular dependence of the resonance field $H_0$ of uniaxial symmetry, as depicted in the upper panel of figure 7 b. The shift in the peak position exceeds 4 kOe when the sample is rotated from $H \parallel c$ to $H \perp c$ orientation. Such an angular dependence is typical for a sheet-like ensemble of randomly oriented magnetic particles, interacting through dipolar forces. The crystalline and shape anisotropies of individual particles lead then to the observed considerable line broadening [41].

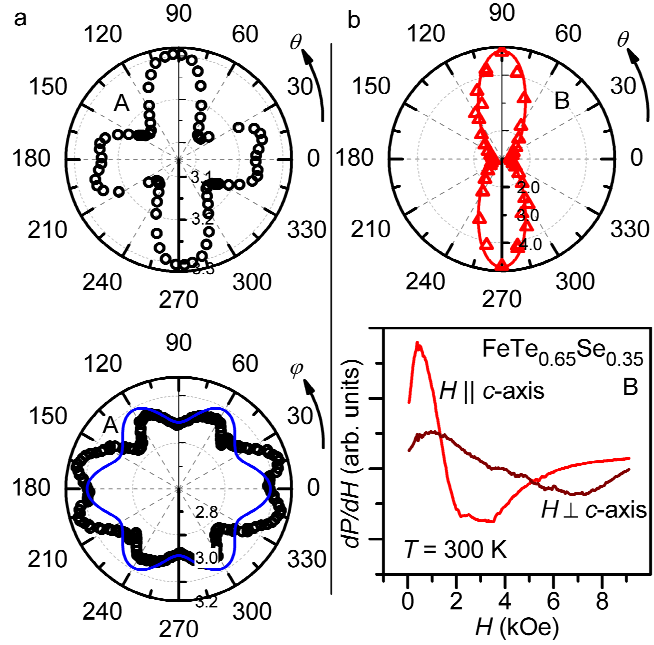

**Figure 7.** (a) The out-of-plane (upper panel) and in-plane (lower panel) angular dependence of FMR line position for line (2) (of Figure 6) in the sample A. The solid line on the in-plane (azimuthal) dependence represents the best fit by the model accounting the hexagonal symmetry of the magnetocrystalline anisotropy with 6-fold axis of symmetry directed along the *c*-axis. (b) The FMR spectrum of the crystal B for two magnetic field orientations (lower panel) and the resonance field dependence on the polar angle (upper panel). The polar angle $\theta$ is measured from the tetragonal crystal axis *c*, while the azimuthal angle $\varphi$ is measured in the *c*-plane from axis *a*.

The magnetic resonance spectrum changes with sample temperature and the line intensity disappears almost completely below 115 K. That temperature dependence is consistent with earlier studies [37,38]. Figure 8 shows the integral intensity for the samples A and B as a function of temperature. The intensity shows a pronounced peak at approx. 125 K and drops rapidly at lower temperatures. The remnant spectrum contains only weak lines between 115 K and 15 K with negligible temperature dependence. Finally, all spectral features disappear completely below 10 K where the samples become superconducting. In particular, we do not see any traces of low-field, low-temperature line observed for K$_{0.8}$Fe$_2$Se$_{1.4}$S$_{0.4}$ [38]. The inset in figure 8 shows temperature dependence of the shift of the resonance line (2) position from the value measured at room temperature. Also this parameter changes rapidly with temperature, in agreement with reports cited earlier.

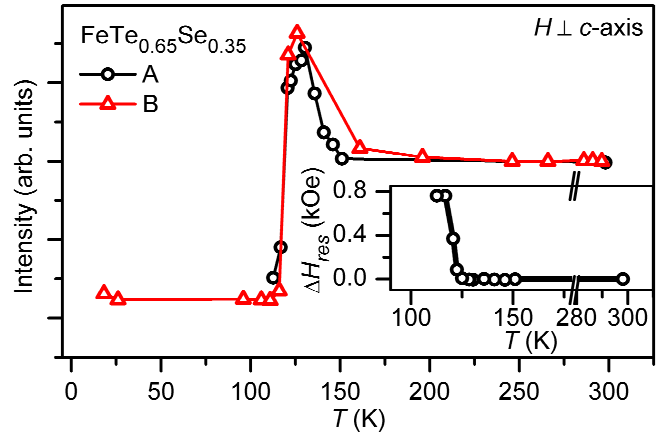

**Figure 8.** Temperature dependence of the integral intensity of the resonance line (2) in the sample A and the sum of lines (3) and (4) in the sample B. The inset shows the temperature dependent shift of the resonance line (2) position in the sample A from the room temperature value. The observed peak in the intensity, just above the structural transition at 100 – 120 K, known as Vervey [29] transition in magnetite, is a characteristic peculiarity for both the magnetite and Fe$_7$(Te,Se)$_8$ phases. The peak position correlates with a fast shift of resonance line position.

Such a behaviour of the MR spectrum is characteristic of a magnetic phase transition and it is usually attributed to a Verwey transition in magnetite [30,42] but it is also typical for the magnetic transition corresponding to the abrupt change in the easy axis of magnetization in ferromagnetic Fe$_7$Se$_8$ in the temperature range of 120 – 130 K [28].

Previous structure, magnetic, transport and optical properties results have indicated a coexistence of magnetic and superconducting phases in iron "11" superconductors [1,2,43-46]. Such a multi-phase system is discussed either as a coexistence of two order parameters on the atomic scale, as combination of incommensurate magnetic order and superconductivity, or as a nanoscale segregation into magnetic and superconducting domains [13]. For the present discussion, we shall focus on the later situation as it is more directly related to our data. There is also a growing evidence that nanometre scale phase separation is an intrinsic feature of "11" Fe superconductors [13,43-45]. In particular, Bahtia *et al.* [44] observed that polycrystalline FeTe$_{0.5}$Se$_{0.5}$ samples "with considerable inhomogeneity through the presence of phase separation and lattice strain, actually possess higher superconducting volume fraction", than homogeneous samples. Hu *et al.* [43], showed nanoscale phase separation and chemical inhomogeneity for several compositions of superconducting FeTe$_x$Se$_{1-x}$. Our systematic investigations of the structure and properties of these magnetic phases and of their influence on the superconductivity show that there is a correlation of nanometre scale structure disorder and related chemical composition fluctuations with superconducting properties in FeTe$_{0.65}$Se$_{0.35}$. The TEM images, shown in figure 2, display structural features in both crystals, demonstrating their similarities and differences. In particular, both crystals contain inclusions of Fe$_3$O$_4$ impurity phases as well as hexagonal symmetry nanometre scale regions, identified as Fe$_7$(Te,Se)$_8$. These phases are also clearly visible in the temperature dependence of the magnetization, shown in figure 5, with almost identical temperature dependence for both crystals. The data collected with FMR spectroscopy confirm such an assignment, supported with a prominent Verwey transition [29,30] in Fe$_3$O$_4$ [34] and magnetic phase transition in Fe$_7$(Te,Se)$_8$, observed at approximately 125 K and with a characteristic angular dependence of the MR spectra as shown in figure 7 a. That assignment is also corroborated by Mössbauer measurements on similar "11" crystals [17,20]. Scanning transmission electron microscopy (STEM) Z-contrast images in figure 3 show the main nanometre-scale structural difference between the crystal A, characterized by data in figure 4 as good bulk superconductor and the crystal B which shows much broader transition and considerably less pronounced specific heat anomaly at $T_c$. Both images show distinct, long, iron-poor bands (white stripes in the figure). There are more bands in the crystal A, they are broader and more diffuse. In addition, the whole background in the image of the crystal A exhibits a fine, well developed pattern with a typical composition fluctuation scale size of ~ 10 nm. Such a pattern is remarkably absent in the crystal B. There, the background is rather smooth indicating a good crystal quality. Also visual inspection of cleaved A and B crystals shows that cleaved surface of the crystal B is shine metallic-like and that of faster cooled crystal A is tarnished. It is remarkable, that the crystal A demonstrates much more disorder than the crystal B and at the same time it is a much better superconductor.

Our image of Z-contrast background in figure 3 a show a distinct nanometre-scale disorder pattern originating from non-equilibrium crystal growth during fast cooling of the liquidus-solidus interface. The mechanism, how nanoscale disorder and related to it chemical fluctuations of crystal composition could enhance superconductivity, is not clear yet. Our STEM data in figure 3 (left panel) look practically identical to that of HAADF pattern recently observed by Hu et al [43] identified by these authors with EELS analysis, as an inhomogeneity in Te/Se contents in FeTe$_{0.7}$Se$_{0.3}$. However, for the present we shall focus on the importance of the iron-deficient hexagonal symmetry regions observed in our crystals.

Appearance of iron-deficient white bands (see, figure 3) can be caused by the development of Fe$_7$Se$_8$-like regions during growth of Fe(Te,Se) crystals. Such regions are completely underdeveloped, despite of their very large concentration in the fast grown crystals and, therefore, they are less pronounced and diffused. On the other hand, the bands are much narrower and better defined in the crystal grown slowly. The difference in superconducting properties between crystals grown with various speed can be then explained assuming that inhomogeneities, in particular their boundaries, and/or stripes are necessary to enhance superconductivity [47-50]. In that case, the existing boundaries between tetragonal host matrix and hexagonal iron deficient regions may play a role of such inhomogeneities. Therefore, the superconducting state properties are improved for the crystals with larger number of tetragonal/hexagonal phase boundaries. There are several models, discussed in the recent literature, which include scenarios of chemical disorder and composition fluctuations, in particular rapid-quench-induced disorder [51], phase separation and magnetic nanodomains as ingredients of enhancement of superconductivity. Our data on crystalline disorder and coexistence of magnetic regions and superconductivity do demonstrate importance of these ingredients but do not allow us, at present, to distinguish a clear mechanism. In particular, magnetic, hexagonal phase nanodomain regions are also a source of crystalline and chemical disorder. More site-specific (low temperature scanning tunnelling microscopy (LT STM) and atomic force microscopy (AFM)) studies are needed to elucidate the spin and charge distribution in nanodomains in superconducting state and to resolve the issue of coexistence of magnetism and superconductivity in iron chalcogenides. On the other hand, as already observed in [44,52], the very presence of disorder (phase separation) might be more important than the details of its structure.

## 5. Conclusions

In both studied FeTe$_{0.65}$Se$_{0.35}$ crystals we observe nanostructures of different symmetry and inhomogeneity of chemical composition. Experimental data demonstrate the presence of nanometre scale hexagonal regions coexisting with tetragonal host lattice, a chemical disorder demonstrating nonhomogeneous distribution of ions in the crystal lattice and hundreds-of-nanometres-long iron-deficient bands. We attribute hexagonal regions to magnetic phases of FeSe, in particular to Fe$_7$(Te,Se)$_8$. Structural differences between studied crystals can be partially attributed to the different growth rate, as discussed above. Our results suggest that inhomogeneous distribution of host atoms might be an intrinsic feature of superconducting FeTeSe chalcogenides. It is in agreement with earlier EELS analysis pointing importance of inhomogeneous anion distribution. Evidence of a surprising correlation was found, that a faster grown crystal of inferior crystallographic properties is a better superconductor.

Summarizing, in FeTeSe the chemical disorder originating from kinetics of crystal growth process influences superconducting properties. In particular, our data support an observation, that ions inhomogeneous spatial distribution and small inclusions of hexagonal phase chalcogenides with nanoscale phase separation seems to enhance the superconductivity

**Acknowledgements**

We would like to thank J. Fink-Finowicki, V. Domukhovski, R. Diduszko, M. Kozłowski and A. Szewczyk for experimental support. This work was supported in part by the EC through the FunDMS Advanced Grant of the European Research Council (FP7 "Ideas"), by the European Regional Development Fund within the Innovative Economy Operational Programme 2007-2013 No POIG.02.01-00-14-032/08 and by National Science Centre (Poland) based on decision No. DEC-2011/01/B/ST3/02374. We thank T. Dietl for suggesting this research and valuable discussions.

**Table 1.** Summary of the chemical composition, structural parameters and critical temperatures for the examined single crystals of FeTe$_{0.65}$Se$_{0.35}$. The critical temperature $T_c^{onset}$ and the width of transition (90% – 10% criterion) was determined from the measurements of AC magnetic susceptibility.

| Sample | A | B |
|---|---|---|
| **Growth velocity (mm/h)** | ~ 8.0 | ~ 1.2 |
| **Starting composition** | FeTe$_{0.65}$Se$_{0.35}$ | FeTe$_{0.65}$Se$_{0.35}$ |
| **Average composition by SEM/EDX** | Fe$_{0.99}$Te$_{0.66}$Se$_{0.34}$ | Fe$_{0.99}$Te$_{0.67}$Se$_{0.33}$ |
| $\Delta\omega$ **(arc min)** | 6.00 | 1.67 |
| $T_c$ **(K)** | ~ 12.9 | ~ 12.9 |
| $\Delta T_c$ **(K)** | ~ 2 (narrow) | ~ 6.8 (wide) |
| $a$ **(Å)** | 3.8036 | 3.8020 |
| $c$ **(Å)** | 6.0921 | 6.0937 |
| **-4$\pi\chi'$ at 2 K** | 1.82 * | < 1 ** |
| **Cleavage** | weak (bulk) | strong (layer) |
| **Additional Bragg peaks (2$\theta$ degree, Cu K$\alpha$ radiation) (degree)** | A$_0$: 32.02; 39.48; 48.84; 54.32; A$_1$: 32.02; 49.60; | B$_0$: 25.86; 29.74; B$_1$: 31.74; 42.28; 48.66; 59.26 |

* As determined with the external field value of 1 Oe; it is equal 1 after correction for demagnetization field.
** The susceptibility does not saturate even at 2 K, almost 10 K below the onset of $T_c$.

**Table 2.** Lorentzian line parameters obtained from the fit of the MR spectra shown in figure 6 with equations (1) and (2).

| Parameter | *I* (arb. unit) | *ΔH* (Oe) | *H$_0$* (Oe) |
|---|---|---|---|
| **Line 1** | 551 ± 12 | 3032 ± 39 | 1158 ± 11 |
| **Line 2** | 1806 ± 17 | 3300 ± 15 | 3319 ± 4 |
| **Line 3** | 2172 ± 29 | 2622 ± 11 | 986 ± 2 |
| **Line 4** | 10590 ± 51 | 5927 ± 14 | 2577 ± 8 |